\documentclass[aps,notitlepage,amsmath,amssymb]{revtex4-1}
\usepackage{graphicx}
\pdfoutput=1

\begin{document}

\title{Scaling properties of generalized two-dimensional Kuramoto-Sivashinsky
  equations}

\author{V. Juknevi\v{c}ius}

\affiliation{Institute of Theoretical Physics and Astronomy, Vilnius University,
A.~Go\v{s}tauto 12, LT-01108 Vilnius, Lithuania}

\begin{abstract}
This paper presents numerical results for the two-dimensional isotropic
Kuramoto-Sivashinsky equation (KSE) with an additional nonlinear term and a
single independent parameter. Surfaces generated by this equation exhibit a
certain dependence of the average saturated roughness on the system size that
indicates power-law shape of the surface spectrum for small wave numbers. This
leads to a conclusion that although cellular surface patterns of definite scale
dominate in the range of short distances, there are also scale-free long-range
height variations present in the large systems. The dependence of the spectral
exponent on the equation parameter gives some insight into the scaling behavior
for large systems.
\end{abstract}
\maketitle

\section{Introduction}

The Kuramoto-Sivashinsky equation (KSE) in its dimensionless form for some
field $h$ can be written as
\cite{article:sivashinsky_1979,article:kuramoto_tsuzuki_1976,article:procaccia_et.al_1992,article:jayaprakash_et.al_1993}
\begin{equation}
	\partial_t h = -\nabla^2 h - \nabla^4 h + (\nabla h)^2\,.
	\label{eq:KSE}
\end{equation}
This equation stands as a paradigmatic model for chaotic spatially extended
systems and can be used to study the connections between chaotic dynamics at
small scales and apparent stochastic behavior at large scales.  It is an
example of an extended, deterministic dynamical system that exhibits complex
spatio-temporal phenomena.  The KSE has been derived for the purpose of
describing the intrinsic instabilities in laminar flame fronts
\cite{article:sivashinsky_1979} and phase-dynamics in reaction-diffusion
systems \cite{article:kuramoto_tsuzuki_1976}.  The equation (\ref{eq:KSE}) in
one- and two-dimensional cases has been a subject of active research for about
three decades, and its scaling properties have even been an object of some
controversy \cite{article:procaccia_et.al_1992,article:jayaprakash_et.al_1993}.

This paper presents some results obtained from a less researched
\emph{generalized} version of KSE.  There have been many different
generalizations of the KSE used for different purposes. Some of them involve
adding damping terms to (\ref{eq:KSE})
\cite{article:sivashinsky_1979,article:paniconi_et.al_1997}, some introduce
spatial anisotropy \cite{article:krug_et.al_1995} or additive random noise
\cite{article:lauritsen_et.al_1996}. 


In the case presented here, there is an additional nonlinear term
$\nabla^2(\nabla h)^2$ introduced to the KSE (\ref{eq:KSE}).  The
two-dimensional generalized Kuramoto-Sivashinsky equation of this form with an
additive Gaussian white noise has been used as a model equation for amorphous
solid surface growth
\cite{article:linz_et.al_exp_2000,article:linz_et.al_2001}. The equation in
this model originally has five parameters that are needed in order to reproduce
the experimental data in the simulations and to examine the correspondence with
microscopic properties of the surface growth process
\cite{article:linz_et.al_exp_2000}.  However, for theoretical investigations of
the long-time and large-scale behavior of the system, the noise term can be
neglected and the remaining deterministic equation can be rescaled into the
dimensionless form with only one independent parameter $\alpha$:
\begin{equation}
	\partial_t h = -\nabla^2 h - \nabla^4 h- \alpha \nabla^2(\nabla h)^2 + (\nabla h)^2\,.
	\label{eq:GKSE}
\end{equation}
Equation (\ref{eq:GKSE}) has also been used as a model for nano-scale pattern
formation induced by ion beam sputtering
\cite{{article:cuerno_et.al_2006},{article:cuerno_et.al_2011}}.

The two-dimensional generalized Kuramoto-Sivashinsky equation (GKSE)
(\ref{eq:GKSE}) in the context of this paper describes the evolution of a
(2+1)-dimensional surface, i.e., a surface which is defined as a function on a
two-dimensional plane and is growing in the direction perpendicular to that
plane.  The surface profile $h(\boldsymbol{r},t)$ is defined as the surface
height $h$ at the position $\boldsymbol{r}=(x,y)$ on the square $[0,L]^2$ of
size $L$ in the plane $\mathbb{R}^2$ at time $t$ or, more
generally, as a function:
\begin{equation}
	h:\quad [0,L]^2 \times \mathbb{R}^{+}\rightarrow\mathbb{R}\,.
\end{equation}

We solve the equation (\ref{eq:GKSE}) for different values of the parameter
$\alpha$ using the \emph{finite difference} method with periodic boundary
conditions, the time step $\Delta t=0.005$, spatial discretization step $\Delta
x =0.71086127010534\,$. Such a seemingly bizarre number for the discretization
step $\Delta x$ is actually a good approximation of the value that is needed in
order for the system with periodic boundary conditions to be able to contain
ordered patterns that appear in some other versions of the generalized KSE. The
equation is solved for system sizes $L$ ranging from about 45 to about 1000
(i.e., on the $N \times N$ lattices with $N$ from 63 to 1400, where
$L=N\,\Delta x$) and in two cases up to about 1422 ($N=2000$). The methods of
numerical solution for (\ref{eq:GKSE}) are presented and compared in
\cite{article:linz_et.al_numerik_2002}.

\section{Kinetics of the surface roughness}

The surface \emph{roughness} $w(t)$, also called the \emph{surface width}, is
one of the most important characteristics of a surface \cite{book:barabasi}. It
is defined as the standard deviation, or, synonymously, root mean square (rms)
deviation, of the surface height  $h(\boldsymbol{r},t)$ from its average value
$\bar{h}(t):=\langle h(\boldsymbol{r},t) \rangle_{\boldsymbol{r}}$, at some
time $t$:
\begin{equation}
	w(t):=\sqrt{\Big \langle \big( h(\boldsymbol{r},t)-\bar{h}(t) \big) ^2 \Big\rangle_{\boldsymbol{r}}}\,.
	\label{eq:rough}
\end{equation}
Here and throughout the whole paper we denote the averaging by $\langle \cdots
\rangle$ with some subscript that shows over what entities the averaging takes
place.  $\langle \cdots \rangle_{\boldsymbol{r}}$ denotes spatial average over
the whole surface, $\langle \cdots \rangle_{t}$ temporal average, $\langle
\cdots \rangle_{\eta}$ ensemble average, $\langle \cdots
\rangle_{|\boldsymbol{r}|=r}$ spatial average over all $\boldsymbol{r}$ whose
length is $r$ etc.

\begin{figure}
	{\centering\includegraphics[width=0.5\textwidth]{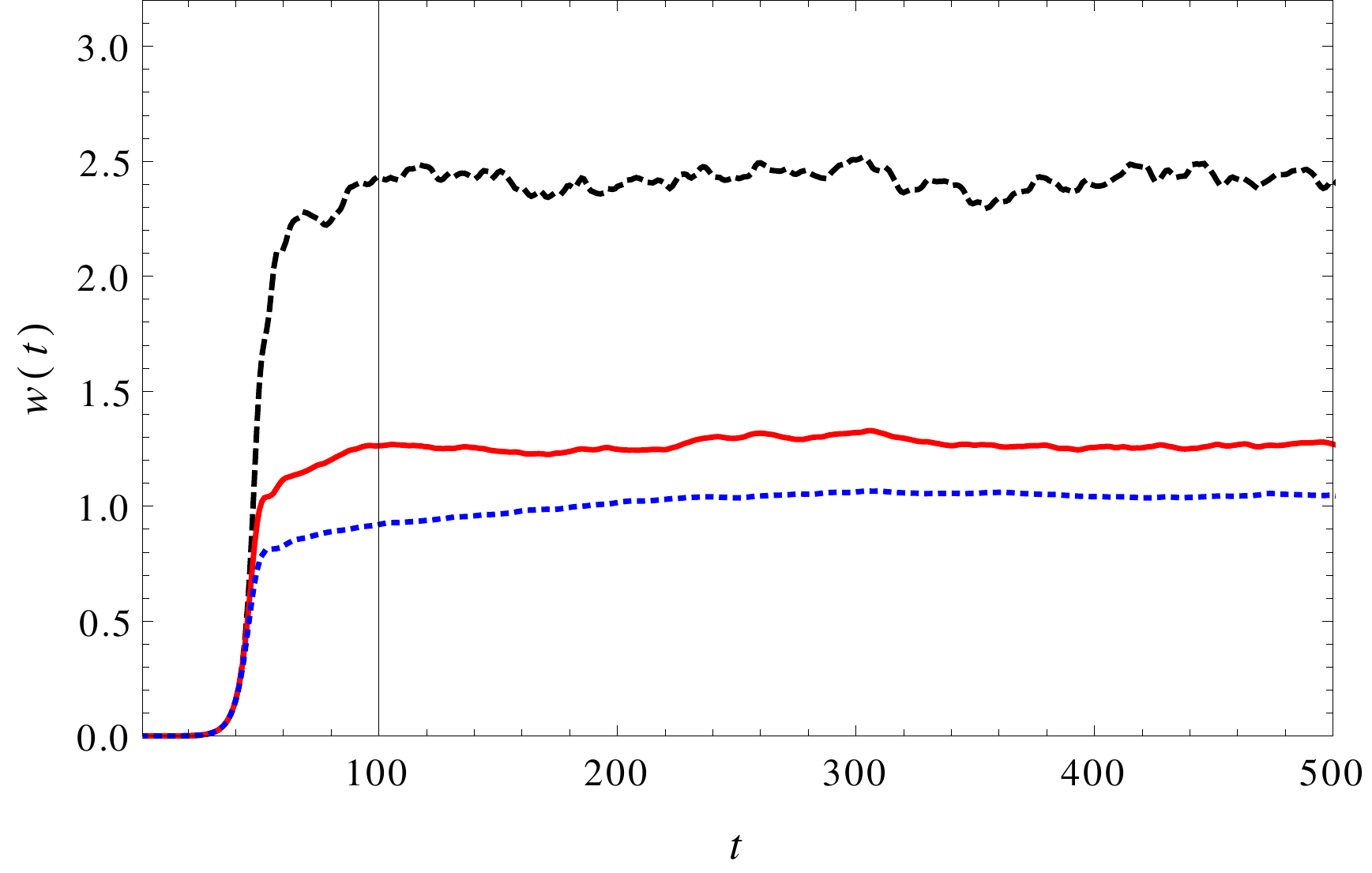}\par}
  \caption{Kinetics of the surface roughness $w(t)$ for short times ($t<500$)
    for surfaces with initial roughness $w(0)=10^{-4}$ evolving according to
    (\ref{eq:GKSE}) with parameter values $\alpha=0$ (\emph{black, dashed
      line}), $\alpha=0.5$ (\emph{red, continuous line}), $\alpha=1$
    (\emph{blue, dotted line}). System size $L\approx 355$ (in lattice units,
    $N=500$).}
	\label{fig:w-saturation}
\end{figure}

For smaller values of parameter $\alpha$ (say, $\alpha<5$) the kinetics of
$w(t)$ due to the evolution of the surface resulting from (\ref{eq:GKSE}) seems
to follow a distinct pattern. Starting from a random surface with some small
initial roughness $w(t=0)\ll 1$, the roughness begins to grow almost
exponentially, but at some time $t_{\times}\approx 100$ this growth slows down
and later on crosses over to a stationary regime where it oscillates about its
average (\emph{saturation}) value $w_{\mathrm{sat}}$ (see Fig.~\ref{fig:w-saturation}).

In the saturation regime, the dynamics of $w(t)$ becomes a statistically
stationary process with time independent average and other statistical
characteristics that are the same for different realizations (different
realizations differ in the initial surface profile, as the evolution equation
itself is deterministic). However, in order to consider the long time behavior
(after saturation) of a statistically stationary process, it is important to take
time averages over sufficiently long time intervals, having in mind that the
minimal required averaging time must be at least several times longer than the 
\emph{correlation time} of the process, defined as the time lag value at which 
the normalized autocorrelation function of $w(t)$ effectively falls to zero.
This correlation time might strongly depend on the parameter $\alpha$ and 
the system size $L$.

In the stationary regime the roughness is chaotically oscillating about some
average value which we denote $w_{\mathrm{sat}}$ and call \emph{saturated
roughness}. This value is calculated as the time average of $w(t)$ in the
stationary regime and is virtually the same when averaged over different time
intervals of the stationary regime (given that these intervals are sufficiently
long) and for different realizations.

This saturated surface roughness can be theoretically defined as a time average
over an time interval of length $T$ that goes to infinity, starting from the
time $t_0$ where the saturation regime is surely reached:
\begin{equation}
	w_{\mathrm{sat}}=\lim_{T\rightarrow\infty}\big\langle w(t) \big\rangle_{t\in[t_{0},\,t_{0}+T)}\,.
	\label{eq:wsat}
\end{equation}

Our numerical investigation shows that, for system sizes $N$ and parameter
values $\alpha$ considered in this paper, it is sufficient to take $t_0\geq
3000$ and $T\geq 10000$ to get the saturation values $w_{\mathrm{sat}}$ that
differ less than $5\%$ for different realizations. For most of the results
presented here, we have used the values $w_{\mathrm{sat}}$ obtained using
$t_0\geq 9000$ and $T=11000$ and, furthermore, we used the averaged values of
several (from 3 to 10) realizations:
\begin{equation}
	w:=\Big\langle\big\langle w(t) \big\rangle_{t\in[9\cdot 10^{3},2\cdot 10^{4}]}\Big\rangle_{\eta}\,.
	\label{w-estimate}
\end{equation}
In this paper we are interested in the surface patterns produced by GKSE
(\ref{eq:GKSE}) and the dependence of the saturated surface roughness on
parameter $\alpha$ and system size $L$. 
		
\section{Resulting surfaces}

\begin{figure}
  {\centering\includegraphics[width=0.45\textwidth]{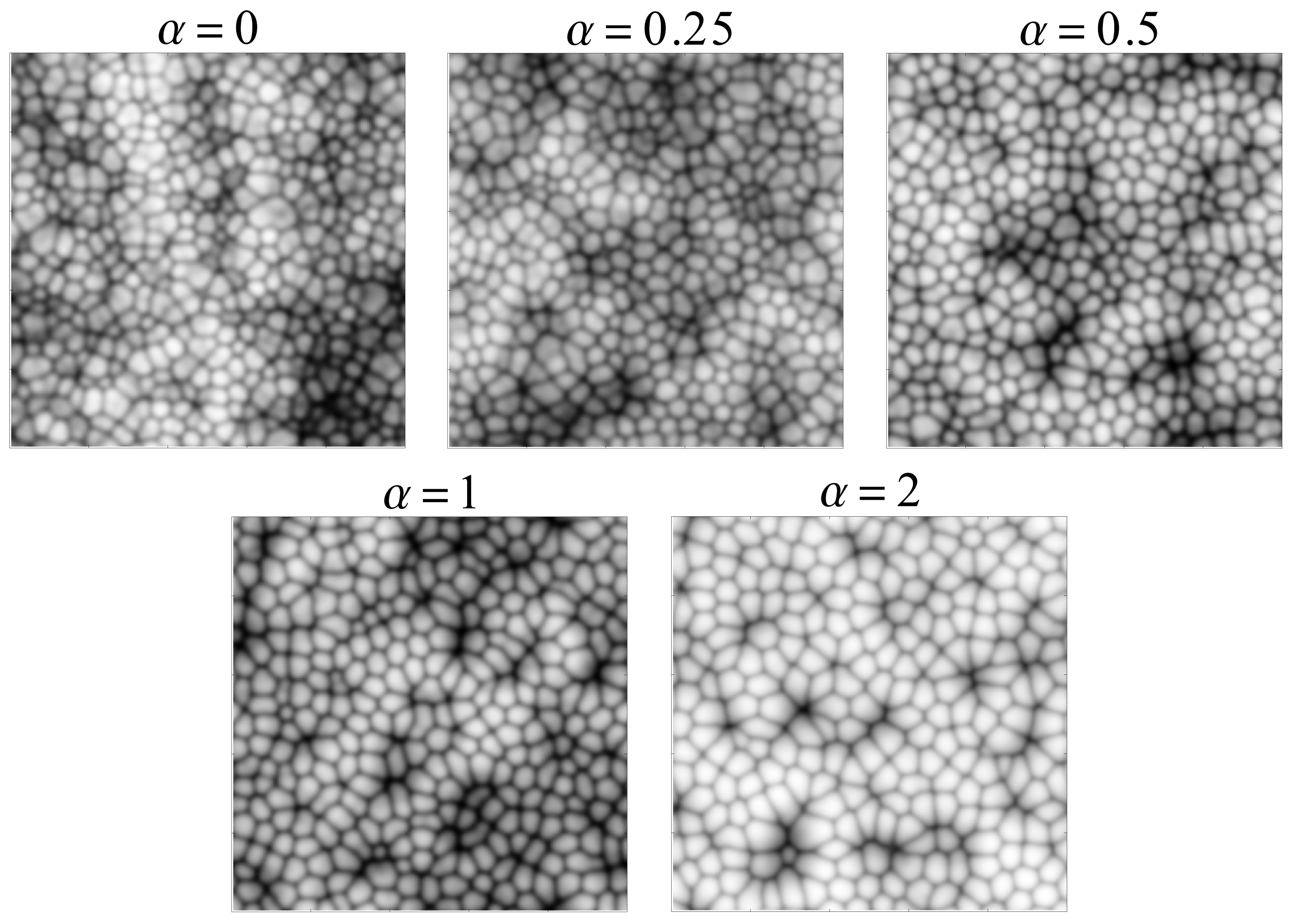}\includegraphics[width=0.45\textwidth]{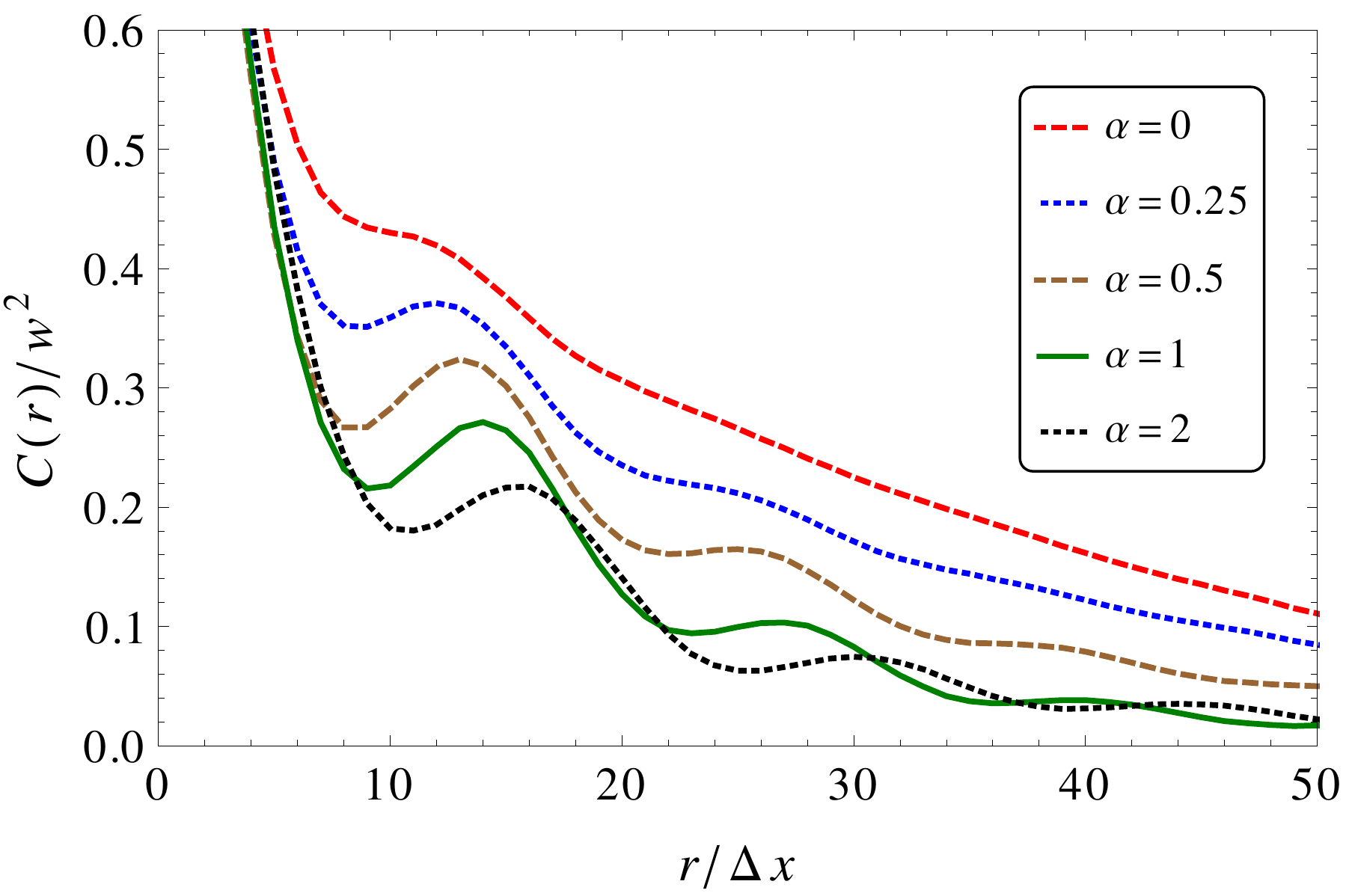}\par}
  \caption{Left panel: Surfaces $h(\boldsymbol{r},t)$ (values of $h$
    coded in gray-scale) evolving according to (\ref{eq:GKSE}) at system size
    $N=250$ ($L\approx178$)  with  parameters $\alpha=0,\,0.25,\,0.5,\,1,\,2$
    at time $t=5000$ (in the stationary regime). Right panel: Normalized
    autocorrelation functions $C(r)$ as defined in (\ref{eq:height-corr}) of
    the surfaces that are shown on the left panel.}
  \label{fig:surf_smallscale}
\end{figure}

\begin{figure}
	{\centering\includegraphics[width=0.5\textwidth]{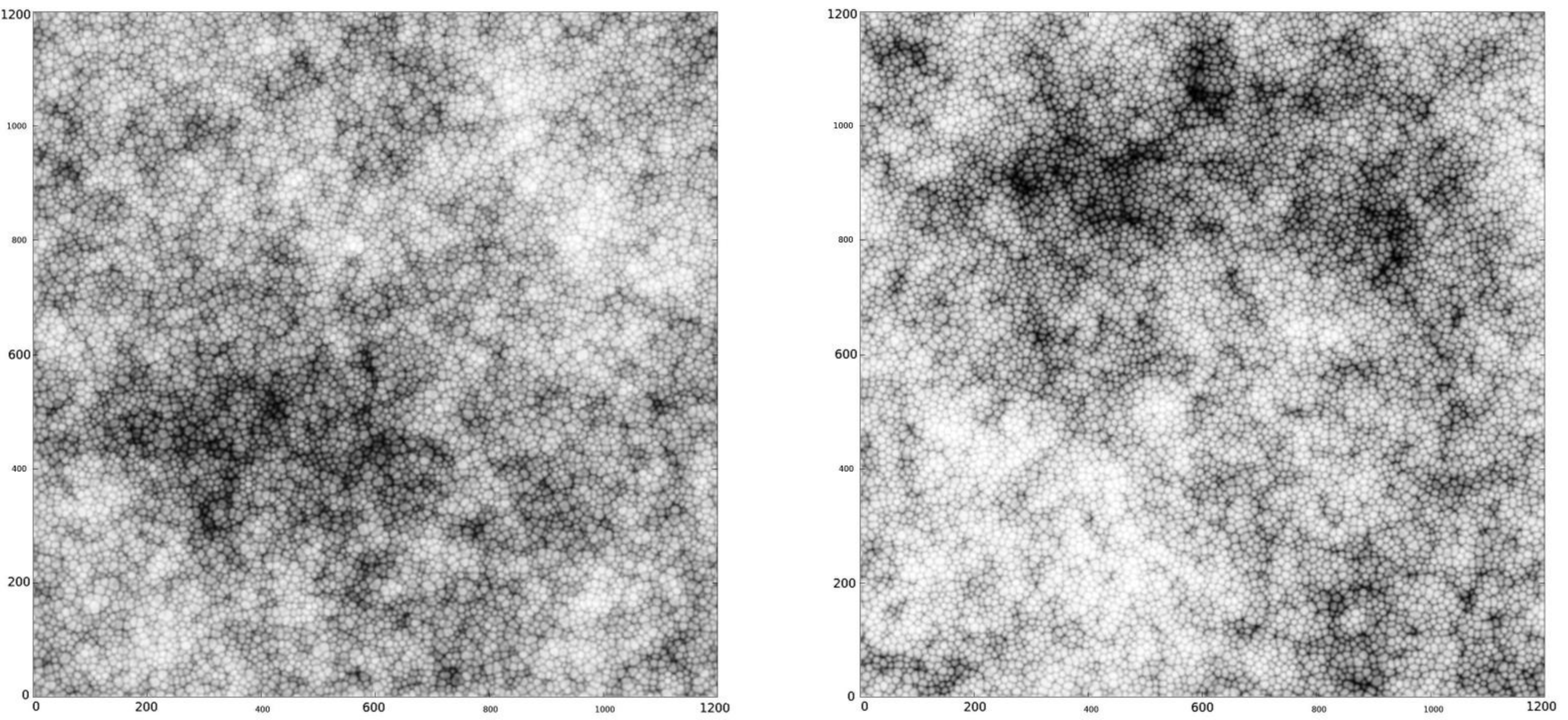}\par}
  \caption{Surfaces (values of $h$ coded in gray-scale) at system size $N=1200$
    ($L\approx853$) evolving according to (\ref{eq:GKSE}) with  parameters
    $\alpha=0$ (left panel) and $\alpha=0.25$ (right panel) at time $t=6000$
    (in the stationary regime).}
	\label{fig:surf1200-01}
\end{figure}

\begin{figure}
  {\centering\includegraphics[width=0.5\textwidth]{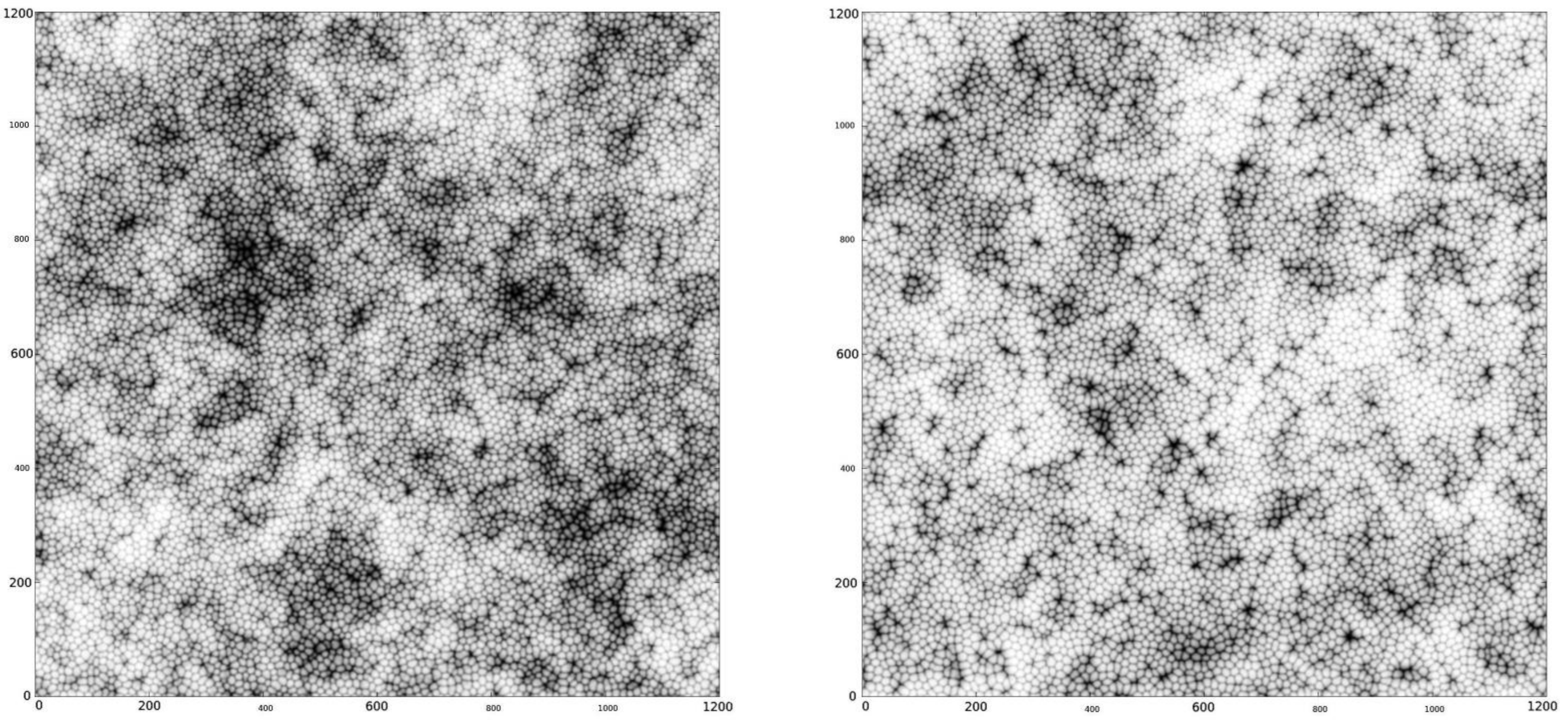}\par}
  \caption{Surfaces (values of $h$ coded in gray-scale) at system size $N=1200$
    ($L\approx853$) evolving according to (\ref{eq:GKSE}) with parameters
    $\alpha=0.5$ (left panel) and $\alpha=1$ (right panel) at time $t=6000$ (in
    the stationary regime).}
  \label{fig:surf1200-02}
\end{figure}

The surfaces, generated by Eq.~(\ref{eq:GKSE}) in the stationary regime have a
disordered cellular structure with cells whose sizes are in a quite narrow
interval (see Figs.~\ref{fig:surf_smallscale}--\ref{fig:surf1200-02}). The
resulting patterns seem to have similar appearance for different system sizes
$L$ (for sizes that are at least several times larger than the typical cell
diameter).

The usual way to investigate the surface patterns is by calculating the surface
height correlation function $C(\boldsymbol{r})$, which is the two-dimensional
autocorrelation function of the surface height $h(\boldsymbol{r})$:
\begin{equation}
	C(\boldsymbol{r})=\big\langle (h(\boldsymbol{r'})-\bar{h})\, (h(\boldsymbol{r'}+\boldsymbol{r})-\bar{h}) \big\rangle_{\boldsymbol{r'}}\,.
	\label{eq:height-corr}
\end{equation}
Since the equation (\ref{eq:GKSE}) is isotropic, the correlation functions
(\ref{eq:height-corr}) $C(\boldsymbol{r})$ of the resulting surfaces should
statistically be independent of the direction of $\boldsymbol{r}$ and, thus,
depend only on its absolute value $r=|\boldsymbol{r}|$. We therefore define the
isotropic height correlation function $C(r)$ as averaged over all directions of
$\boldsymbol{r}$:
\begin{equation}
	C(r)=\Big\langle \big\langle  (h(\boldsymbol{r'})-\bar{h})\, (h(\boldsymbol{r'}+\boldsymbol{r})-\bar{h}) \big\rangle_{\boldsymbol{r'}} \Big\rangle_{|\boldsymbol{r}|=r}\,.
	\label{eq:height-corr-iso}
\end{equation}


Fig.~\ref{fig:surf_smallscale} shows resulting surface patterns for system size
$L\approx178$ (or, in lattice units, $N=250$) and different values of parameter
$\alpha$ in (\ref{eq:GKSE}) and the corresponding normalized height correlation
functions (\ref{eq:height-corr-iso}).  We see that for the surfaces generated
by (\ref{eq:GKSE}) with $\alpha=0$, corresponding to the KSE (\ref{eq:KSE})
case, the height correlation function (\ref{eq:height-corr-iso}) has no maximum
(right panel of Fig.  \ref{fig:surf_smallscale}), just an area of slower decay
at distances $r$ corresponding to the approximate sizes of cells in the
pattern. For $\alpha>0$, the height correlation function obtains a maximum
whose distance corresponds to the average size of cells. We see that this the
distance of this maximum increases with $\alpha$, meaning that the cell size
increases as the parameter $\alpha$ is increased.

Another thing that can be seen in Fig.~\ref{fig:surf_smallscale} is that the
normalized correlation functions $C(r)/w^2$ decrease slowly for small values of
$\alpha\geq 0$ and faster for larger values. That is the first indication of
the influence of parameter $\alpha$ on long-range height correlations.


Large-scale height variations in surfaces produced by Eq.~(\ref{eq:GKSE})
become more distinct as the system size is chosen to be many times larger than
the typical cell size. The examples for some values of $\alpha$ can be seen in
Fig.~\ref{fig:surf1200-01} and Fig.~\ref{fig:surf1200-02}. 

By repeating the simulations with different system sizes $L$, one notices a
slight dependence of the average saturation value of the surface roughness
$w_{\mathrm{sat}}$ on the system size. Indeed, the results show that the
resulting roughness increases when the system size is increased. This shows
that the resulting surface profile contains spatial Fourier components of ever
smaller wave number $k$ (and correspondingly larger wavelength $\lambda$).
Although for small scales (large wave numbers) the structures of definite size
occur, for larger distances (small wave numbers) we get the height variations
with long-range dependence, and (as will be shown in the next two sections)
this dependence has a scale free character.

\section{Power-law surface spectra and scaling of roughness}

This section presents some general considerations about the spatial
power-spectral densities (PSD) of surfaces, their connection to the surface
roughness $w$, and the effects introduced by the finite system size. The
scaling behavior of $w$ when the PSD has a power-law shape is derived. These
theoretical results are compared to the numerically calculated scaling
properties of $w$ for the surfaces generated by (\ref{eq:GKSE}) in the next
section.

\subsection{Surface PSD and roughness}

A two-dimensional surface $h(\boldsymbol{r})$ is a single valued function on
the plane $\boldsymbol{r}=(x,y)\in \mathbb{R}^2$. In order to avoid the zero
frequency component in the spectrum, we calculate the spectrum of the surface
profile $h_c(\boldsymbol{r})$ with zero mean:
\begin{equation}
 	h_c(\boldsymbol{r}):=h(\boldsymbol{r})-\bar{h}\,,
 	\label{eq:hc}
\end{equation}
where $\bar{h}\equiv \langle h(\boldsymbol{r}) \rangle_{\boldsymbol{r}}$ is the
average height of the surface.  Fourier transformation $H(\boldsymbol{k})$ of
this 'centered' surface profile
$h_c(\boldsymbol{r})$:
\begin{equation}
 	H(\boldsymbol{k})=\int\mathrm{d}^2\boldsymbol{r} \,h_c(\boldsymbol{r}) \,\exp(-\mathrm{i} \boldsymbol{k} \cdot \boldsymbol{r})\,.
 	\label{eq:h_FT}
\end{equation}
Here $\boldsymbol{k}=(k_x,k_y)^T$ is the wave vector of spatial Fourier
components of the surface profile.

The surface power-spectral density (PSD) is then defined as:
\begin{equation}
	S_{\boldsymbol{k}}= \frac{1}{L^2} |H(\boldsymbol{k})|^2\,,
	\label{eq:h_PSD}
\end{equation}
where $L$ is the size of the segment of the surface analyzed, i.e.,
$\boldsymbol{r}\in [0,L]^2$.  The integral of the power spectrum
$S_{\boldsymbol{k}}$ over all $\boldsymbol{k}$ is equal to the variance of the
surface height $\sigma_{h}^2$ which by our definition is is equal to the square
of the surface roughness $w^2$:
\begin{equation}
 	\frac{1}{(2\pi)^2}\int\mathrm{d}^2\boldsymbol{k}\,S_{\boldsymbol{k}}=\sigma_{h}^2\equiv w^2\,.
 	\label{eq:PSD_int}
\end{equation}
Since our model is isotropic, $S_{\boldsymbol{k}}$ must depend only on the
absolute value $|\boldsymbol{k}|$, and we can get the one-dimensional power
spectrum $S(k)$ of the surface by integrating the two-dimensional power spectrum over all
wave vectors $\boldsymbol{k}$ of the same absolute value $k$. This can be done
by expressing the wave vector $\boldsymbol{k}$ in the polar coordinates
$\boldsymbol{k}=(k,\phi)$, and then integrating $S_{\boldsymbol{k}}$ over 
all angles $\phi$:
\begin{equation}
	\frac{1}{(2\pi)^2}\int \mathrm{d}^2\boldsymbol{k}\,S_{\boldsymbol{k}}=
	\frac{1}{(2\pi)^2}\int \mathrm{d}k\,k \int_{0}^{2\pi} \!\!\mathrm{d}\phi \,S_{\boldsymbol{k}}=
	\frac{1}{2\pi}\int \mathrm{d}k\,S(k)\,.
	\label{eq:1d-2d-PSD_def}
\end{equation}
From this we get that the one-dimensional PSD of the surface can be expressed
as:
\begin{equation}
	S(k)=\frac{k}{L^2} \, \big\langle |H(\boldsymbol{k})|^2\big\rangle_{|\boldsymbol{k}|=k} 
	=\frac{k}{L^2} \,  |H(\boldsymbol{k})|^2 =k\,S_{\boldsymbol{k}}\,\,.
	\label{eq:1d-PSD_def}
\end{equation}
In Eq.~(\ref{eq:1d-PSD_def}) we have used the fact that for isotropic surfaces 
$\langle |H(\boldsymbol{k})|^2\rangle_{|\boldsymbol{k}|=k}=|H(\boldsymbol{k})|^2$. 
The integral of this one-dimensional PSD $S(k)$ over all wave numbers $k$ also
equals to the square of surface roughness:
\begin{equation}
	\frac{1}{2\pi}\int \!\mathrm{d}k\,S(k)= w^2
	\,\,.
	\label{eq:1d-PSD_int}
\end{equation}

\subsection{Effects of finite system size}

The surfaces in numerical simulations are represented on a $(N\times N)$ matrix
that sets limits to the smallest and the largest possible wave numbers
$k_{\mathrm{min}}$ and $k_{\mathrm{max}}$ that can fit into the system, thus,
'filtering' the theoretically defined PSD $S(k)$. If $\Delta x$ is the spatial
step size in the simulation, then the minimal distinguishable wavelength
$r_{\mathrm{min}}$ in the system is approximately equal to this discretization
step doubled:
\begin{equation}
	r_{\mathrm{min}}\approx 2 \Delta x \,,
\end{equation} 
and maximal wavelength that can fit into the system is of about double system
size:
\begin{equation}
	r_{\mathrm{max}}\approx 2 L= 2N  \Delta x \,.
\end{equation}
Since the distance $r$ corresponds to the wave number $k=\frac{2 \pi}{r}$, we
get the minimal and maximal wave numbers for the system:
\begin{eqnarray}
	k_{\mathrm{min}}\approx&\frac{2 \pi}{r_{\mathrm{max}}}=\frac{\pi}{N \Delta x}\,,\nonumber\\
	\label{eq:minmax_wavenumber}\\
	k_{\mathrm{max}}\approx&\frac{2 \pi}{r_{\mathrm{min}}}=\frac{\pi}{ \Delta x}\,.\nonumber
\end{eqnarray}
The square of the numerically calculated surface roughness, expressed according
to (\ref{eq:1d-PSD_int}) should then be
\begin{equation}
	w^2\approx \frac{1}{2\pi}\int_{k_{\mathrm{min}}}^{k_{\mathrm{max}}}\! \!\mathrm{d}k\,S(k)
	\,\,.
	\label{eq:1d-PSD_int-num}
\end{equation}

If we keep the discretization step $\Delta x$ constant, assuming that surface
patterns for systems of different sizes (up to the smallest wave numbers
allowed by the system size) are statistically the same, then, by increasing the
system size $N$, according to (\ref{eq:minmax_wavenumber}), we reduce the
minimal wavenumber $k_{\mathrm{min}}$ in (\ref{eq:1d-PSD_int-num}). Therefore
the calculated surface roughness must grow with the system size and its scaling
behavior $w(N)$ when $N$ is increased should be able to give us information
about the shape of the surface PSD $S(k)$ for small wave numbers $k\rightarrow
0$.

\subsection{Power-law spatial spectrum for small wave numbers}

Let us assume that the one-dimensional spatial PSD $S(k)$ (\ref{eq:1d-PSD_def})
of a surface has a power-law dependence on $k$ for small wave numbers, smaller
than some wavelength $k_{\mathrm{s}}$:
\begin{equation}
	S(k)=\left\{
	\begin{array}{ccc}
		C\,k^{-\gamma} & \mathrm{for} & k < k_{\mathrm{s}}\,, \\
		S_1(k) & \mathrm{for} & k \geq  k_{\mathrm{s}}\,.\\
	\end{array}
	\right.
	\label{spectrum_pwl}
\end{equation}
Here $C$ is some constant, $S_1(k)$ is the shape of the PSD spectrum for high
wave numbers $k$ that does not interest us, since we are interested in large
scale behavior of the system and, furthermore, assume that $S_1(k)$ doesn't
change when the system size is increased. We also assume that the power for
small wave numbers $\gamma\leq 1$.
 
Then, for a system of finite size and
$k_{\mathrm{min}}<k_{\mathrm{s}}<k_{\mathrm{max}}$, the calculated square of
the roughness is expressed by:
\begin{equation}
	w^2\approx \int_{k_{\mathrm{min}}}^{k_{\mathrm{max}}}\! \!\mathrm{d}k\,S(k)=
	C\,\int_{k_{\mathrm{min}}}^{k_{\mathrm{s}}}\! \!\mathrm{d}k\,k^{-\gamma}
	+ A \,.
	\label{wsquare00}
\end{equation}
Since $k_{\mathrm{s}}$ and $\gamma$ come from the model and $k_{\mathrm{max}}$
comes from numerical scheme, the only variable is $k_{\mathrm{min}}$ which is
inversely proportional to the system size. Therefore the second term in
(\ref{wsquare00}) is just a constant which we denoted by $A$:
\begin{equation}
	A=\int_{k_{\mathrm{s}}}^{k_{\mathrm{max}}}\! \!\mathrm{d}k\,S_1(k)\,.
\end{equation}

There are two qualitatively distinct cases. One is $\gamma=1$ which would
result in infinite $w^2$ for a system of infinite size and another case is
$\gamma<1$ for which even a system of infinite size would have a finite height
variance $w^2$.
For $\gamma=1$, the square of the surface roughness $w$ from (\ref{wsquare00})
grows linearly with the logarithm of the system size $N$:
\begin{equation}
		w^2\approx C \ln \bigg( \frac{k_{\mathrm{s}}}{k_{\mathrm{min}}}\bigg)+A
		= C \ln \big( N \big)+B \,.
		\label{wsquare_gamma1}
\end{equation}
Here $C$ and $B=A-\ln(N_{\mathrm{s}})$ with
\begin{equation}
		N_{\mathrm{s}}=\frac{\pi}{k_{\mathrm{s}} \Delta x}
\end{equation}
being constant. The last expression in (\ref{wsquare_gamma1}) is more useful,
since we don't know the exact value of $N_{\mathrm{s}}$.  We see that when
$\gamma=1$, the surface roughness $w$ goes to infinity for infinite size system
($N \rightarrow \infty$). This means that in this case the influence of
long-range height variations grows with the system size.
	
When $\gamma<1$	, from (\ref{wsquare00}) we get the following scaling relation:	
\begin{equation}
		w^2\approx \frac{C}{1-\gamma}\Big( k_{\mathrm{s}}^{1-\gamma} - k_{\mathrm{min}}^{1-\gamma} \Big) +A=
		D \Big(1-\big(N/N_{\mathrm{s}}\big)^{-(1-\gamma)} \Big)+A
\end{equation}
with constants $A$, as defined above, and $D$, which defined as
\begin{equation}
		D=\frac{C}{1-\gamma}\,\Big(\frac{\pi}{N_{\mathrm{s}} \Delta x}\Big)^{1-\gamma}\,.
\end{equation}
Again, since we don't know the value of $N_{\mathrm{s}}$, we express the
scaling of the surface roughness as	
\begin{equation}
		w^2\approx C_1 - C_2\,N^{-(1-\gamma)}
		\label{wsquare_gamman1}
\end{equation}	
with the constants
\begin{equation}
		C_1=D+A\,,
\end{equation}
and
\begin{equation}
		C_2=D \,N_{\mathrm{s}}^{1-\gamma}\,.
\end{equation}
We see that in this case (for $\gamma<1$) the surface roughness $w$ has a
finite value for a system of infinite size:
\begin{equation}
		w_{\infty}:=\lim_{N \rightarrow \infty} w(N) =\sqrt{C_1}\,.
		\label{roughinfsiz}
\end{equation}

\section{Numerical results}

In this section we present our numerical results for surfaces generated by
(\ref{eq:GKSE}): the one-dimensional power-spectral densities (PSD) $S(k)$
(\ref{eq:1d-PSD_def}) calculated from the autocorrelation function $C(r)$
(\ref{eq:height-corr-iso}) for system sizes up to $N=500$ and the scaling of
the roughness $w$ which gives shapes of the $S(k)$, based on the considerations
of the previous section.

\begin{figure}
  {\centering\includegraphics[width=0.5\textwidth]{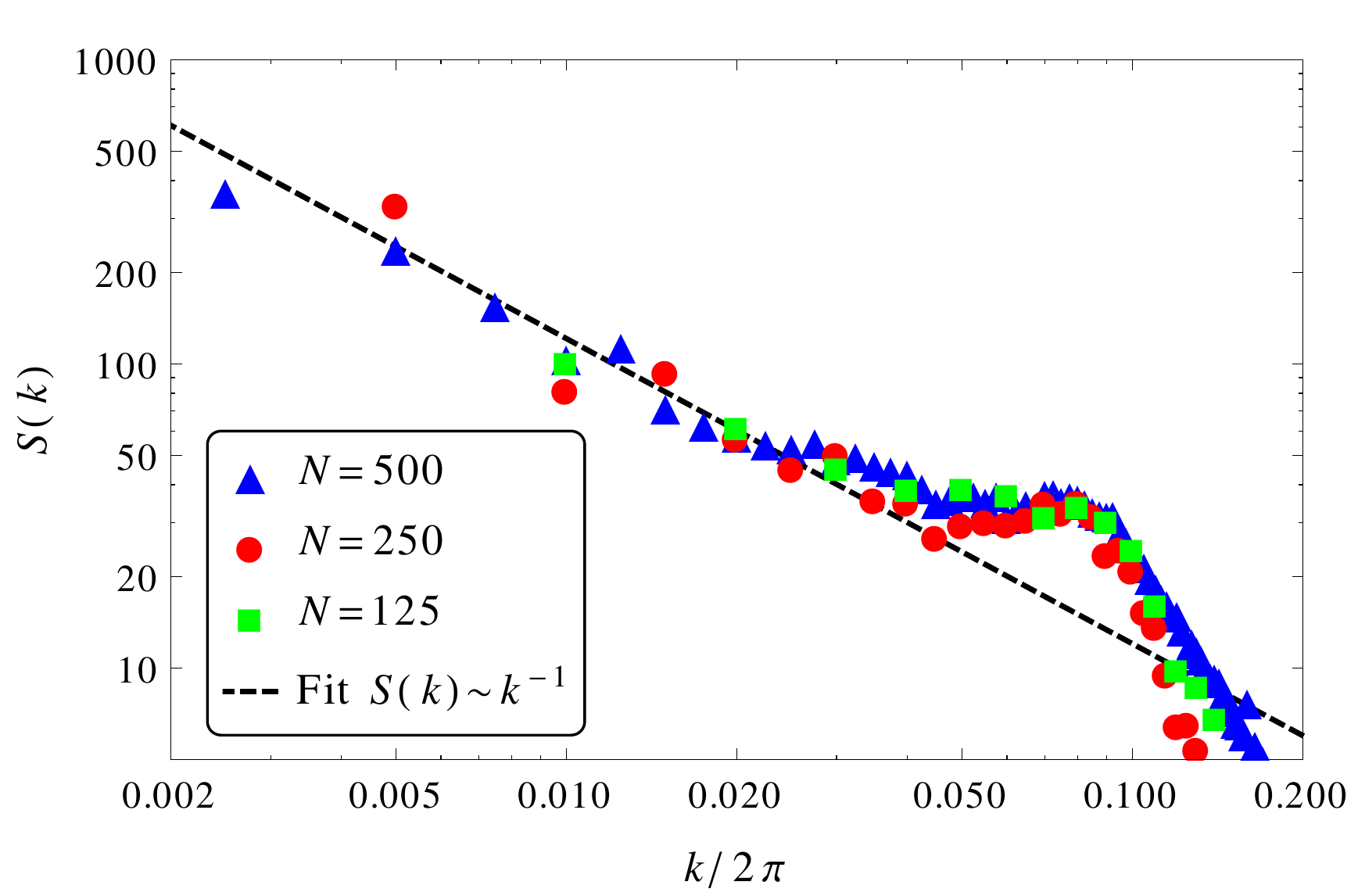}\par}
  \caption{(log-log scale) Numerically calculated PSD of the surfaces at time
    $t=10^4$ produced by (\ref{eq:GKSE}) with parameter $\alpha=0$ for system
    sizes (in lattice units) $N=500$ (blue triangles), $N=250$ (red filled
    circles) and $N=125$ (green squares). The black dashed line represents
    power-law fit with exponent $\gamma=1.003$.}
  \label{fig:specsize0}
\end{figure}	

\subsection{Surface spectra}

\begin{figure}
	{\centering\includegraphics[width=0.5\textwidth]{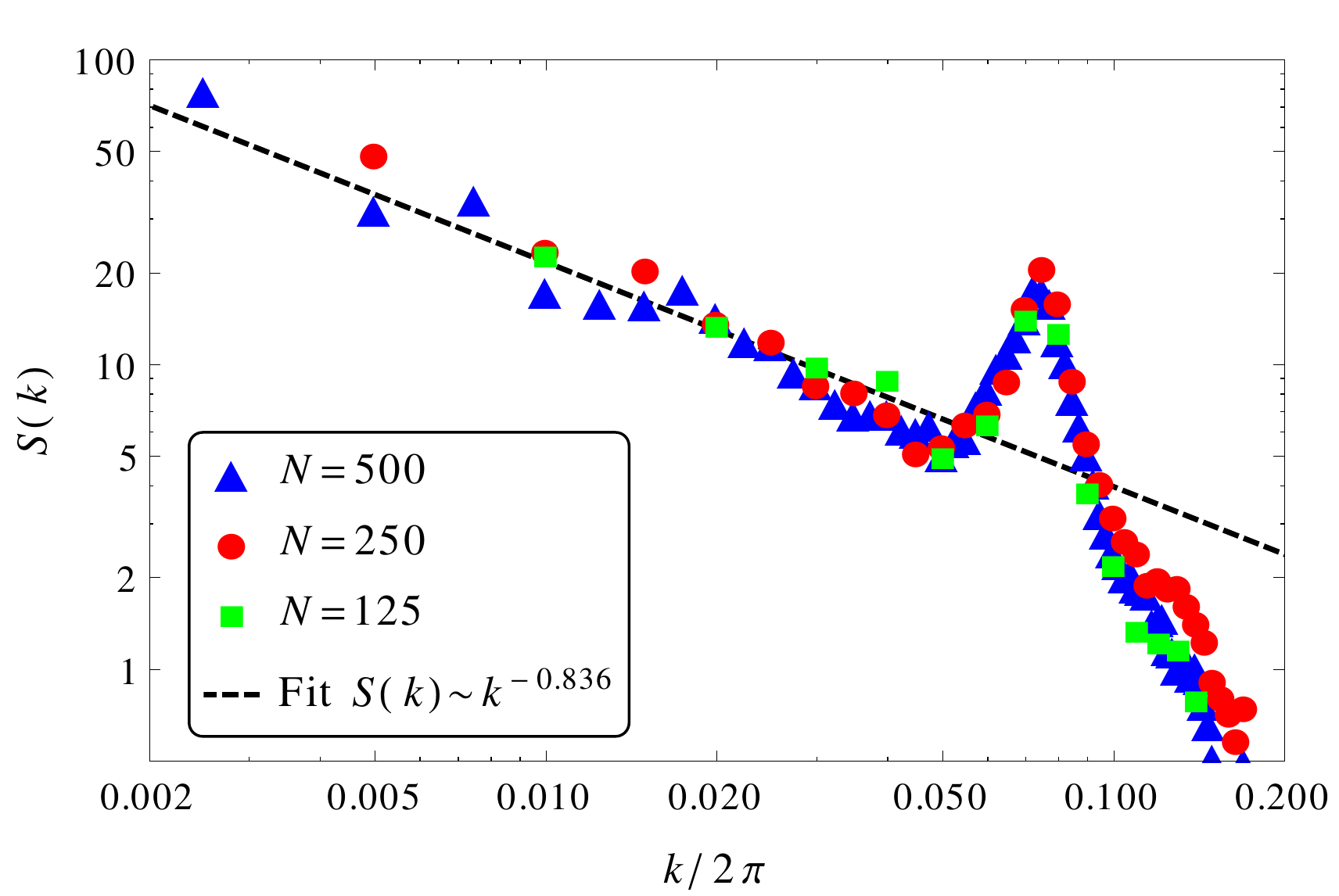}\par}
  \caption{(log-log scale) Numerically calculated PSD of the surfaces at time
    $t=10^4$ produced by (\ref{eq:GKSE}) with parameter $\alpha=1$ for system
    sizes (in lattice units) $N=500$ (blue triangles), $N=250$ (red filled
    circles) and $N=125$ (green squares). The black dashed line represents
    power-law fit with exponent $\gamma=0.836$.}
	\label{fig:specsize1}
\end{figure}	

Since the equation (\ref{eq:GKSE}) is isotropic, the direction of the wave
vector $\boldsymbol{k}$ does not matter in statistical description of height
variations, and in spectral analysis of the surface patterns,  the wave number
$k=|\boldsymbol{k}|$ is sufficient to describe the occurring spatial modes. We
can therefore analyze one-dimensional surface spectra $S(k)$, defined in
(\ref{eq:1d-PSD_def}).

We calculate the one-dimensional surface spectrum $S(k)$ that depends only on
the wave number $k$ using the Wiener-Khinchin theorem which states that the
power-spectral density can be obtained from the Fourier transform of the
autocorrelation function.  Although the height correlation function $C(r)$
(\ref{eq:height-corr-iso}) depends only on the absolute value $r$ of the shift
$\boldsymbol{r}$, it is nevertheless a two-dimensional autocorrelation function
of the surface $h(\boldsymbol{r})$. Thus, by applying the two-dimensional
Fourier transform on the height correlation function
(\ref{eq:height-corr-iso}), we get the two-dimensional PSD (\ref{eq:h_PSD})
from which we calculate the one-dimensional PSD $S(k)$ using
(\ref{eq:1d-PSD_def}).

The numerically calculated height correlation function for an isotropic surface
$h(\boldsymbol{r})$ has been defined in (\ref{eq:height-corr-iso}):
\begin{equation}
	C(r)=\Big\langle \big\langle h_c(\boldsymbol{r'})\, h_c(\boldsymbol{r'}+\boldsymbol{r}) \big\rangle_{\boldsymbol{r'}} \Big\rangle_{|\boldsymbol{r}|=r}\,.
	\nonumber
\end{equation}
Here, as before, $ h_c(\boldsymbol{r})= h(\boldsymbol{r})-\bar{h}$ is the
surface height with its average value subtracted, the so-called 'centered'
surface profile.  The two-dimensional PSD $S_{\boldsymbol{k}}$ is calculated
according to (\ref{eq:h_PSD}) with (\ref{eq:h_FT}):
\begin{equation}
	S_{\boldsymbol{k}}=\frac{1}{L^{2}}\int \mathrm{d}^2 \boldsymbol{r}'\,\int\mathrm{d}^2 \boldsymbol{r}''\,
	h_{c}(\boldsymbol{r}')\,h_{c}(\boldsymbol{r}'')
	\mathrm{e}^{-\mathrm{i}\boldsymbol{k}\cdot(\boldsymbol{r}''-\boldsymbol{r}')}\,.
	\label{eq:2d-PSD_full}
\end{equation} 
By changing the variable
$\boldsymbol{r}''\rightarrow\boldsymbol{r}'+\boldsymbol{r}$ and switching the
order of integration, we get
\begin{equation}
	S_{\boldsymbol{k}}=\int\mathrm{d}^2 \boldsymbol{r}\,\mathrm{e}^{-\mathrm{i}\boldsymbol{k}\cdot\boldsymbol{r}}
	\frac{1}{L^{2}}\int \mathrm{d}^2 \boldsymbol{r}'\,
	h_{c}(\boldsymbol{r}')\,h_{c}(\boldsymbol{r}'+\boldsymbol{r})\,,
	\label{eq:2D-PSD-expl}
\end{equation}
where, according to (\ref{eq:height-corr}),
\begin{equation}
	\frac{1}{L^{2}}\int \mathrm{d}^2 \boldsymbol{r}'\,
	h_{c}(\boldsymbol{r}')\,h_{c}(\boldsymbol{r}'+\boldsymbol{r})
	\equiv 	 
	\big\langle h_c(\boldsymbol{r'})\, h_c(\boldsymbol{r'}+\boldsymbol{r}) \big\rangle_{\boldsymbol{r'}} 
	= C(\boldsymbol{r})
	\label{eq:acorr_calc}
\end{equation}
is the height correlation function of the surface $h(\boldsymbol{r})$.

Since the surfaces $h(\boldsymbol{r})$ generated by (\ref{eq:GKSE}) are
statistically isotropic, the correlation function (\ref{eq:acorr_calc})
effectively depends only on $r=|\boldsymbol{r}|$ and can be denoted by $C(r)$.
Using this, expressing the two-dimensional integration in
(\ref{eq:2D-PSD-expl}) over $\boldsymbol{r}$ in the polar coordinates
$\boldsymbol{r}=(r,\phi)$ and integrating over the angular part, we get
\begin{equation}
	S_{\boldsymbol{k}}=\int \mathrm{d}r\, r \,C(r)\, \int_{0}^{2\pi} \!\mathrm{d}\phi \,
	 \mathrm{e}^{\mathrm{i}kr\cos \phi}\,.
\end{equation}
Substituting this into (\ref{eq:1d-2d-PSD_def}) gives the one-dimensional PSD
of the surface:
\begin{equation}
	S(k)=k\,2\pi\, \int\!\mathrm{d}r\,r\,C(r)\, J_0(kr)\,,
	\label{eq:suspec-corr}
\end{equation}
Here $J_0(kr)$ is the Bessel function of the 1st kind:
\begin{equation}
	 J_0(kr)=\frac{1}{2\pi}\int_{0}^{2\pi}\!\mathrm{d}\phi\, \mathrm{e}^{\mathrm{i}kr\cos \phi}\,.
\end{equation}

Figures \ref{fig:specsize0} and \ref{fig:specsize1} show the calculated surface
spectra of the surfaces generated by (\ref{eq:GKSE}) with parameter values
$\alpha=0$ (the KSE case) and $\alpha=1$, respectively. In each figure, the
spectra for systems of sizes (in lattice units) $N=125$, $N=250$ and $N=500$
are shown. Each spectrum is obtained by averaging over 6 different
realizations. The figures confirm the above assumption that increasing the
system size does not change the small scale structures, since the spectra at
large values of $k$ coincide. The appearance of low wave number modes at larger
system sizes and the approximate power-law behavior of the spectrum $S(k)$ can
also be seen.

Of course, for larger systems, the numerical calculation of the two-dimensional
autocorrelation function (\ref{eq:height-corr}) and surface spectrum
(\ref{eq:suspec-corr}) directly can take a very long time. Therefore, the
possibility to obtain $S(k)$ from scaling of the surface roughness is very
useful.

\begin{figure}		
  {\centering\includegraphics[width=0.5\textwidth]{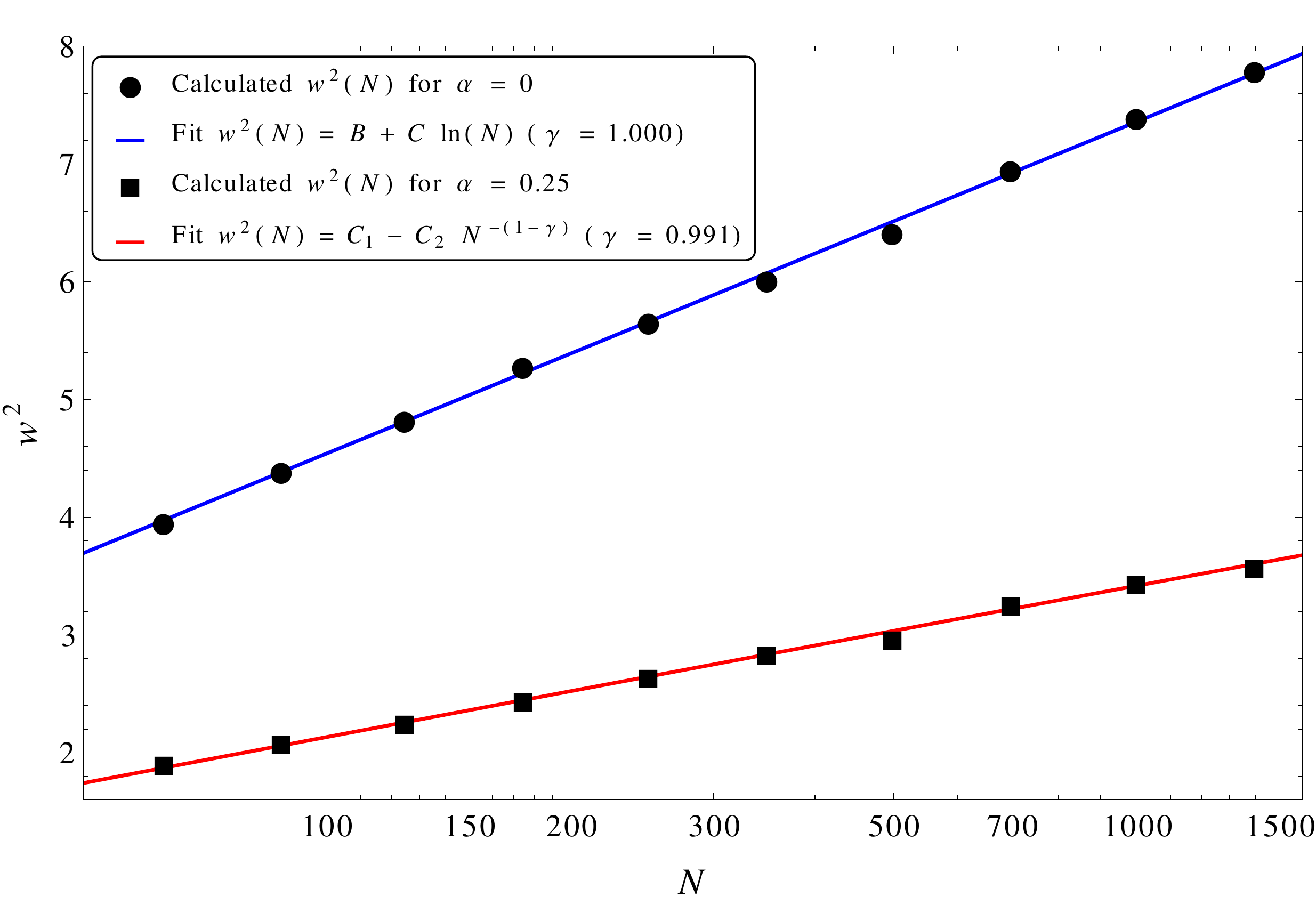}\par}
  \caption{(log-linear scale) Black symbols: numerical results for surfaces
    evolving according to (\ref{eq:GKSE}) with $\alpha=0$ (filled circles) and
    $\alpha=0.25$ (filled squares). Time averaged square of the surface
    roughness $w^2$ plotted as a function of the system size $N$ (in lattice
    units). Lines: fits of the results for $\alpha=0$ (blue line) and
    $\alpha=0.25$ (red line) by (\ref{wsquare_gamma1}) and
    (\ref{wsquare_gamman1}) respectively, corresponding to the scaling surface
    spectrum (\ref{spectrum_pwl}). Resulting fit parameters for $\alpha=0.0$:
    $B=-1.0935$, $C=1.224$; for $\alpha=0.25$:  $\gamma=0.991$, $C_1=67.770$,
    $C_2=68.284$ . }
  \label{fig:scaling01}
\end{figure}

\subsection{Scaling of the surface roughness}
 
\begin{figure}
	{\centering\includegraphics[width=0.5\textwidth]{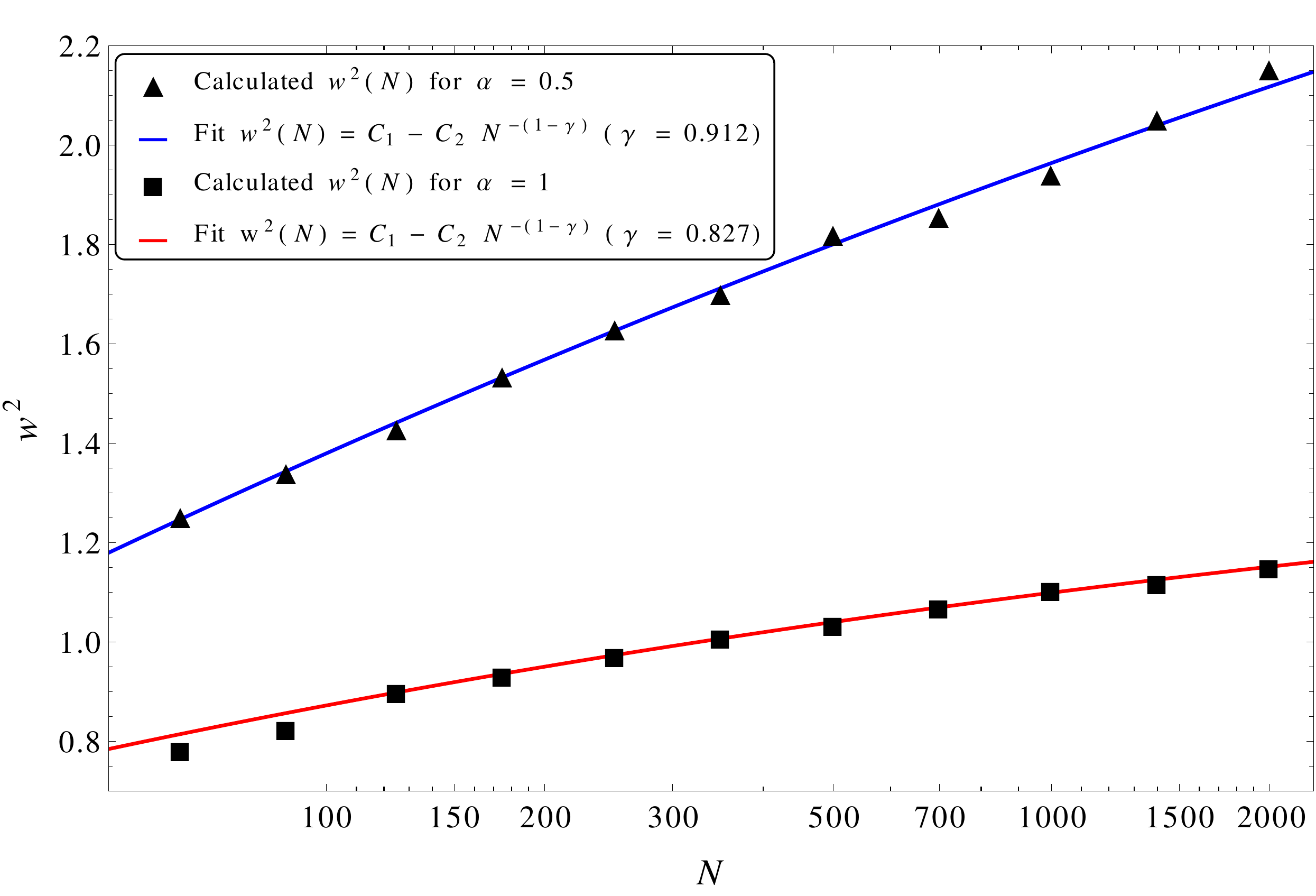}\par}
  \caption{(log-linear scale) Black symbols: numerical results for surfaces
    evolving according to (\ref{eq:GKSE}) with $\alpha=0.5$ (filled triangles)
    and $\alpha=1.0$ (filled squares).Time averaged square of the surface
    roughness $w^2$ plotted as a function of the system size $N$ (in lattice
    units). Lines: fits of the results for $\alpha=0.5$ (blue line) and
    $\alpha=1.0$ (red line) by (\ref{wsquare_gamman1}) corresponding to the
    scaling surface spectrum (\ref{spectrum_pwl}). Resulting fit parameters for
    $\alpha=0.5$:  $\gamma=0.912$, $C_1=4.554$, $C_2=4.768$; for $\alpha=1.0$:
    $\gamma=0.827$, $C_1=1.563$, $C_2=1.532$ . }
	\label{fig:scaling02}
\end{figure}	

We have investigated the dependence of this saturated surface roughness on the
size of the model system for different values of the parameter $\alpha$ in
(\ref{eq:GKSE}).  We investigated how the square of $w$ estimated by
(\ref{w-estimate}) depends on the system size $N$ (in lattice units) for sizes
$N=63$, $N=88$, $N=125$, $N=175$, $N=250$, $N=350$, $N=500$, $N=700$, $N=1000$,
$N=1400$, $N=2000$. The results are shown in Fig.~\ref{fig:scaling01} for
$\alpha=0$ and $\alpha=0.25$, Fig.~\ref{fig:scaling02} for $\alpha=0.5$ and
$\alpha=1$.

For $\alpha=0$ (the KSE case), the scaling exponent (as introduced in
(\ref{spectrum_pwl})) $\gamma=1$ fits the results best. Thus the dependence of
$w^2$ on $N$ can be approximated by (\ref{wsquare_gamma1}) and the results
should form a straight line in the \emph{log-linear} plot, and, indeed they do
as can be seen in Fig.~\ref{fig:scaling01} (filled black circles). The blue
line in Fig.~\ref{fig:scaling01} fitted to data for sizes from $N=88$ to
$N=1400$.

For $\alpha=0.25$, the scaling exponent $\gamma\approx 0.991$ fits the results
best and the dependence of $w^2$ on $N$ can be approximated by
(\ref{wsquare_gamman1}). Fig. \ref{fig:scaling01} shows the calculated results
(filled black rectangles). There seems to be a good agreement between the data
and the fitting curve which is fitted to data for sizes from $N=125$ to
$N=1400$ (red line in Fig. \ref{fig:scaling01}).

The results for $\alpha=0.5$ and $\alpha=1$ give scaling exponents are
$\gamma\approx0.912$  and $\gamma\approx0.827$, respectively, and the
dependence of $w^2$ on $N$ can be approximated by (\ref{wsquare_gamman1}).
Fig.~\ref{fig:scaling02} shows the calculated results (black triangles for
$\alpha=0.5$ and black rectangles $\alpha=1$) together with their respective
fits (blue and red lines in Fig.~\ref{fig:scaling02}) .
		
\section{Summary and discussion}

The generalized Kuramoto-Sivashinsky equation (\ref{eq:GKSE}) produces surfaces
with disordered cellular patterns (Figs.~\ref{fig:surf_smallscale},
\ref{fig:surf1200-01} and  \ref{fig:surf1200-02}). The size of the average size
of a cell depends on equation parameter $\alpha$ and constitutes a definite
scale in the surface pattern (see the peaks on the correlation functions in
\ref{fig:surf_smallscale}). However, in larger systems, the long-range height
variations of different character become apparent (Figs.~\ref{fig:surf1200-01}
and \ref{fig:surf1200-02}). We investigate these long-range height variations
for several values of parameter $\alpha$ by calculating the corresponding
scaling relations of the surface roughness.

The square of the surface roughness $w$ (\ref{eq:rough}) which is by definition
equal to the variance of the surface height $h(\boldsymbol{r},t)$, $w^2\equiv
\sigma^2_h$, can also be expressed as an integral (\ref{eq:1d-PSD_int}) of the
one-dimensional power spectral density $S(k)$, given in (\ref{eq:1d-PSD_def}),
of the surface over all wave numbers $k$. Since the finite size $L=N \Delta x$
of the system and the discretization step $\Delta x$ in numerical simulations
define the approximate lower and upper cut-off values
(\ref{eq:minmax_wavenumber}) for the possible wave numbers, the estimated value
of $w^2$ (\ref{eq:1d-PSD_int-num}) depends on the system size.  As indicated by
the surface spectra in Figs.~\ref{fig:specsize0} and \ref{fig:specsize1}, for
systems of increasing size, the small-scale patterns (corresponding to higher
wave numbers $k$) remain statistically the same, and, additionally, new lower
wave number modes arise in larger systems. Therefore by increasing the size of
the system, from the corresponding change in the calculated value of $w^2$
(\ref{eq:1d-PSD_int-num}), the shape of the spatial power spectrum $S(k)$ can
be extracted. 

If the behavior of $S(k)$ for small wave numbers $k$ (say when
$k<k_{\mathrm{s}}$, for some $k_{\mathrm{s}}$) follows the inverse power-law
(\ref{spectrum_pwl}), $S(k)\propto k^{-\gamma}$, with $\gamma>0$, this
indicates that long-range, scale-free height variations are present. There can
be qualitatively distinct cases for different values of the exponent $\gamma$.
If $\gamma= 1$, then we get the scaling relation (\ref{wsquare_gamma1}) which
implies that $w$ will grow indefinitely as the system size goes to infinity.
If, on the other hand, $\gamma<1$, then we get the scaling relation
(\ref{wsquare_gamman1}), therefore $w$ will approach finite value
(\ref{roughinfsiz}) as the system size goes to infinity.

For the surfaces generated by numerical simulations of the generalized
isotropic Kuramoto-Sivashinsky equation (\ref{eq:GKSE}), we indeed see the
indications that the spatial power spectral density $S(k)$ follows a power-law
(\ref{spectrum_pwl}), since the theoretically calculated scaling relations
(\ref{wsquare_gamma1}) and (\ref{wsquare_gamman1}) fit the numerically
established scaling of $w^2$ well (see Figs. \ref{fig:scaling01} and
\ref{fig:scaling02}). For the Kuramoto-Sivashinsky equation (\ref{eq:KSE}), we
get the spectral exponent $\gamma=1$ and the scaling relation
(\ref{wsquare_gamma1}). For parameter values $\alpha>0$ the scaling exponent
$\gamma$ decreases resulting in the scaling relation (\ref{wsquare_gamman1})
giving the surfaces of finite roughness as the size of the system goes to
infinity, and thus showing that the generalized Kuramoto-Sivashinsky equation
(\ref{eq:GKSE}) with $\alpha>0$ does not belong to the same universality class
as the KSE (\ref{eq:KSE}).

\end{document}